\documentstyle[multicol,epsfig,aps,prb,array]{revtex}

\newcommand \be{\begin{equation}}
\newcommand \ee{\end{equation}}
\newcommand \ba{\begin{eqnarray}}
\newcommand \ea{\end{eqnarray}}

\begin{document}
\title{Tunneling-assisted impact ionization fronts in semiconductors}

\author{P. Rodin \cite{St.P.}}
\address{
Institute for Theoretical Physics, Technical University Berlin,
Hardenbergstrasse 36, 10623 Berlin, Germany
\\
Centrum voor Wiskunde en Informatica,
P.O.Box 94079, 1090 GB Amsterdam, The Netherlands
}
\author{U. Ebert and W. Hundsdorfer}
\address{
Centrum voor Wiskunde en Informatica,
P.O.Box 94079, 1090 GB Amsterdam, The Netherlands
}
\author{I.V. Grekhov}
\address{
Ioffe Physicotechnical Institute, Politechnicheskaya 26,
194021, St. Petersburg, Russia
}

\date{\today}
\maketitle

%************************************************************************
%------------------------ABSTRACT----------------------------------------

\begin{abstract}

We propose a novel type of ionization front in layered semiconductor
structures. The propagation is due to the interplay of
band-to-band tunneling and impact ionization.
Our numerical simulations show that the front can be triggered
when an extremely sharp voltage ramp ( $\sim 10 \; {\rm kV/ns}$ )
is applied in reverse direction to a Si $p^+-n-n^+-$structure
that is connected  in series with an external load.
The triggering occurs after a delay of 0.7 to 0.8 ns.
The maximal electrical field at the front edge exceeds $10^6 \;
{\rm V/cm}$. The front velocity $v_f$ is 40 times faster than
the saturated drift velocity $v_s$.
The front passes through the $n-$base with a thickness of
$100 \; {\mu m }$ within approximately 30~ps,
filling it with dense electron-hole
plasma. This passage is accompanied by a voltage drop from
8~kV to dozens of volts. In this way a voltage pulse
with a  ramp up to  $500 \, {\rm kV/ns}$  can be applied
to the load. The possibility to form a kilovolt pulse with such
a voltage rise rate sets new frontiers in pulse power electronics.
\end{abstract}

\begin{multicols}{2}

%***************************************************************************
%------------------------------INTRODUCTION---------------------------------

\section{Introduction}

Impact ionization and tunneling (Zener breakdown) are the two
most fundamental mechanisms capable of creating high concentrations
of free carriers in a semiconductor within a short time interval.
The respective threshold electrical fields differ by almost one order
of magnitude, e.g.,  $2 \cdot 10^5 \, {\rm V/cm}$ and $10^6 \, {\rm V/cm}$
for impact and tunneling ionization in Si, respectively \cite{Sze,Shur}.
Avalanche impact ionization can be easily achieved by applying a
sufficiently strong external electrical field.
This process underlies the operation of many semiconductor devices
such as avalanche transistors, IMPATT and TRAPATT diodes, {\it etc.}
\cite{Sze,Shur,Shaw}.
The most interesting scenario corresponds to
the propagation of a {\it superfast ionization front}:
a narrow impact ionization region travels from the cathode
to the anode with a velocity $v_f$ much higher than
the saturated drift velocity $v_s$, leaving a high density plasma behind
\cite{PRA68,DEL70,GRE79,BEN85,ALF87,GRE89}.
In contrast to impact ionization,
direct tunneling of electrons from the valence
to the conductance band is hard to archieve
in the bulk of uniformly doped semiconductor layers:
while the applied external voltage
is being increased,
impact ionization typically sets in first, inducing an avalanche
multiplication of free carriers. This causes the conductivity to increase
and prevents further increase of the applied voltage.
For this reason, tunneling ionization
is generally assumed to be relevant
only in heavily doped  $p-n-$junctions
due to the strong internal electrical fields \cite{Sze}.

In this article we demonstrate that the threshold of tunneling ionization
in the bulk of a Si $p^+-n-n^+-$structure can be reached
under the same experimental conditions as for triggering impact
ionization fronts \cite{GRE79,BEN85,ALF87,GRE89}.
Typically a solitary ionization front is triggered by applying a sharp
voltage ramp ($ \ge 1 \; {\rm kV/ns}$) to the $p^+-n-n^+-$structure
in reverse direction  \cite{GRE79,GRE89,MIN00}.
In structures with kilovolt $p-n-$junctions and
large cross-sections, this process is used for
sharpening electrical pulses \cite{GRE89,FOC97,KAR95,KAR01}.
This technique allows one to
reach voltage ramps of up to $10 \, {\rm kV/ns}$, the state of the
art in modern pulse power electronics.
We demonstrate, that when such a sharp ramp $A \sim \; 10 \; {\rm kV/ns}$
is applied to a fully depleted reversely biased Si $p^+-n-n^+-$structure,
the threshold of tunneling ionization
$\sim 10^6 \, {\rm V/cm}$ is reached
after less than 1~ns, which turns out to be faster
than the initiation of avalanche impact ionization.
The resulting breakdown takes the form of an ionization front
that propagates due to the combined effect of tunneling and impact
ionization. Compared to the traditional impact ionization fronts
in pulse sharpening diodes \cite{GRE79,BEN85,ALF87,GRE89,FOC97,KAR95}
and TRAPATT-diodes \cite{PRA68,DEL70},
these {\it tunneling-assisted impact ionization fronts}
are expected to be much faster
and generate higher plasma concentrations. Their practical
application may set new frontiers in pulse power electronics.

\section{The model}

We consider a Si $p^+-n-n^+-$structure with sharp $p^+-n$ and
$n-n^+-$transitions and the following parameters:
the width of the $n-$base is $W=100 \; {\rm \mu m}$, the cross-section
area is $S = 0.002 \; {\rm cm^2}$, the doping levels are
$N_d \approx 10^{14} \; {\rm cm^{-3}}$ in the
$n-$base and $N_{a,d}^{+} \approx 10^{19} - 10^{20} \; {\rm cm^{-3}}$
in the contact $p^{+}-$ and $n^{+}-$layers, respectively.
These parameters correspond to a typical Si power diode
with a stationary breakdown voltage $\sim 1.5 \; {\rm kV}$
\cite{Sze}. We choose the initial bias of $V_0 = 1 \, {\rm kV}$,
closely below the voltage of stationary avalanche breakdown.
For this bias, the $n-$base is fully depleted from major carriers (electrons)
and equilibrium minor carriers (holes).

The device is connected to a voltage source $V(t)$ in series with
a load resistance $R = 50 \; \Omega$ as sketched in
Fig.\ 1. Since it is convenient to work with a positive electrical
field $E$ for the reverse bias, we put the $n-n^+-$junction
on the l.h.s.\ at $z = 0$ and the $p^+-n-$junction on the r.h.s.\
at $z = W$. The voltage $V(t)$ applied to the structure and the load
is in linear approximation
\be
V(t) = V_0 + A \, t ,
\label{voltage}
\ee
where $A$ is the voltage ramp. Hereinafter we keep $A=10$~kV/ns.
The voltage on the device is denoted as $U(t)$ and related to
$V(t)$ through the Kirchhoff equation $V = U + R I$,
where $I$ is the total current.

We use a minimal model which accounts only for the basic transport
processes, for band-to-band impact
ionization and tunneling ionization.
The continuity equations for
electrons and holes $n$ and $p$ are written in one-dimensional
approximation
\ba
\partial_t n -
\partial_z\left(v_n(|E|)\;n\right)
&=& G(n,p,|E|),
\\
\partial_t p +
\partial_z\left(v_p(|E|)\;p\right)
&=& G(n,p,|E|).
\ea
and complemented by the Poisson equation
\be
\partial_z E = \frac{q}{\varepsilon\varepsilon_0}\;
\Big(p-n+N_d(z)-N_a(z)\Big)
\ee
and the Kirchhoff equation.  We assume drift-dominated transport
and approximate the carrier velocites as
\cite{JAC77}
\ba
v_n(|E|) = v_s \;\frac{|E|}{E_{sn}+|E|},
&~~&
v_p(|E|) = v_s \;\frac{|E|}{E_{sp}+|E|},
\label{velocities2}
\\
\mbox{where }\qquad\qquad
v_s &=& 10^7 \; {\rm cm/s},
\qquad
\\
E_{sn} = 8.0 \cdot 10^3 \; {\rm V/cm},
&~~&
E_{sp} = 2.32 \cdot 10^4 {\rm V/cm}.
\nonumber
\end{eqnarray}
Since the ionization processes are fast
and develop in the bulk of the device,
we solve these equations in the $n-$base only.
The different effect of $p^+-n-$ and $n-n^+-$junctions
on electron and hole concentrations in the $n-$base is modelled
by mixed boundary conditions: $\partial_z n=0$, $p=0$ at $z=0$ and
$\partial_z p=0$, $n=0$ at $z=W$ \cite{boundary}.
The generation term $G(n,p,|E|)$  contains band-to-band tunneling and
impact ionization terms:
\be
G(n, p, |E|) = G_T (|E|) + G_I(n, p, |E|),
\label{generation}
\ee
The tunneling term $G_T (|E|)$ models electron tunneling
from the valence band to the conduction band \cite{tunneling}
\ba
G_T (|E|) = \alpha_T E^2 \, e^{- b_T/|E|}, \\
\nonumber
\alpha_T = \frac{q^2}{3 \pi^2 \hbar^2} \sqrt{\frac{2 m}{E_g}},
\qquad
b_T = \frac{\pi}{4 q \hbar} \sqrt{2 m E_g^3},
\label{tunneling}
\ea
where $q$ and $m$ are electron charge and effective mass,
respectively, $E_g$ is the bandgap,
and $\hbar$ is Planck's constant.
The impact ionization term $G_I(n, p, |E|)$ is chosen as
\begin{eqnarray}
\label{impact_ionization}
G_I(n,p,|E|) &=&
\alpha_{n}(|E|) \, v_n(|E|) \, n \;\;\Theta(n - n_{cut})
\\
&&+\;\alpha_{p}(|E|) \, v_p(|E|)\, p \;\; \Theta(p - p_{cut}),
\nonumber
\end{eqnarray}
\begin{equation}
\alpha_n(|E|) \equiv \alpha_{ns} \, e^{- b_n/|E|},
\;
\alpha_n(|E|) \equiv \alpha_{ps} \, e^{-b_p/|E|},
\label{ion_coef}
\end{equation}
where the impact ionization
coefficients and the characteristic fields are given by \cite{KUZ75}
\begin{eqnarray}
\alpha_{ns} = 7.4 \cdot 10^5 \; {\rm cm^{-1}},
&~&
\alpha_{ps} = 7.25 \cdot 10^5 \; {\rm cm^{-1}},
\nonumber\\%\qquad
b_n = 1.1 \cdot 10^6 \; {\rm V/cm},
&~&
b_p = 2.2 \cdot 10^6 \; {\rm V/cm},
\end{eqnarray}
and $\Theta(x)$ is the step-function.
The cut-offs $n_{cut}$ and $p_{cut}$ have been
introduced in (\ref{impact_ionization}) to mimic the discreteness
of the charge carriers. The purpose is to
exclude unphysical ionization avalanches
initiated by tiny fractions of electrons or holes that can cause
premature triggering of the front in the simulations.
We chose $n_{cut} = p_{cut} = 10^{9} \, {\rm cm^{-3}}$
in the simulations and discuss the effect of this and
other choices in Section V and Fig.\ 6.

Since we shall investigate processes on a sub-nanosecond time scale,
we neglect all types of recombination and thermal generation.
We also assume that the n-base is free of deep-level defects or
parasitic impurities that can assist tunneling or
serve as deep-level electron traps capable to release
electrons or holes in high electric field \cite{ROD01}.
We discuss these model assumptions in detail in Sections IV and V.

We use a uniform space-time grid with the number of points
of the order of several thousands both in time and space.
The spatial discretization is based on a conservative
formulation, in terms of fluxes describing the inflow
and outflow over cells $[x-\Delta x/2, x + \Delta x/2]$,
where  $ \Delta x$ is the grid width in space. Whereas
the diffusive fluxes have been approximated in a standard
fashion \cite{text_book_numerics}  with second order accuracy,
for the convective fluxes a third order upwind biased formula
has been chosen in order to reduce the numerical oscillations.
Time discretization is based on a
second order backward differentiation formula.
The temporal backward differentiation formula gives
an implicit system that is solved at each time step.
For reasons of accuracy the time step $\Delta t$ is chosen
small compared to $\Delta x / v_s$, where $v_s$ is the upper bound of
the drift (convective) velocity, and with such small step size that the
implicit system can be solved by a straightforward functional iteration.
Details on these spatial and temporal discretizations can be found in
Ref.~\onlinecite{VHB}.

\section{Numerical solutions}

The basic features of the numerical solutions
are summarized in Figures 2 and 3. They show
the external characteristics of the transient
and the internal dynamics, respectively.
The voltage on the device first increases
and reaches the maximal value 8~kV,
several times higher than the voltage of stationary
breakdown. During this stage, the electrical field
in the structure increases like
\be
E(x,t) \approx \frac{V(t)}{W} - \frac{bW}{2} + b x, \qquad
b \equiv \frac{q N_d}{\varepsilon \varepsilon_0},
\label{field}
\ee
in the same manner as in the TRAPPAT-diode \cite{Sze,DEL70}.
(Here we assume that the displacement current is small and
hence $V(t) \approx U(t)$).
Though the electrical field $E$ at the right boundary
exceeds the effective threshold of impact
ionization $2 \cdot 10^5 \; {\rm V/cm}$ already at $t > 30 \; {\rm ps}$,
impact ionization does not develop
due to the absence of initial carriers. At $t \approx 720 \; {\rm ps}$
the electrical field at the right boundary $x = W$, near
the $p^+-n-$junction, becomes sufficient for tunneling
of electrons from the valence to the conduction band
(Fig.\ 3(a), curve 1).
At this time, the field is above the threshold of impact
ionization in the whole $n-$base, so impact ionization starts
as soon as electrons and holes are supplied by tunneling.
The rate of impact ionization increases with concentration
and eventually overheads tunneling at the carrier density
$n \sim p \sim 10^{13} \; {\rm cm^{-3}}$ (see Fig.\ 4).
The rapid ionization process near the right boundary
forms an initial nucleus of dense electron-plasma with concentration
$n,p \sim 2 \cdot 10^{17} \, {\rm cm^{-3}}$
that is capable to fully screen the electrical field
(Fig.\ 3(a), curves 2, 3). The screening is accomplished
at $t = 750 \, {\rm ps}$. It is accompanied
by a fast drop of the voltage on the device and an increase of current in
the circuit (Fig.\ 2). This drop corresponds to the first step
in the falling $U(t)$-curve.

Consecutively, the highly conducting plasma region expands to the left
in the form of a superfast ionization front (Fig.\ 3(b)).
The velocity of the front $v_f$ is approximately 40 times faster
than the saturated drift velocity $v_s$ of the individual carriers;
hence the carrier motion is negligible.
The front propagates due to the combined effect
of tunneling and impact ionization, followed by Maxwell relaxation
in the generated plasma and consecutive electric screening.
Snapshoots of the field and concentration profiles together
with respective generation rates at some instant of time
are shown in Fig.\ 5.
Here tunneling generates initial carriers in the high field
region at the edge of the ionization front. These carriers are
multiplied further by impact ionization. A narrow region of strong impact
ionization is localized at the edge of the visible concentration front.
The rate of impact ionization in this region is by many orders
of magnitude higher than the rate of tunneling ionization.
Therefore impact ionization dominates the overall increase of the
concentration. However, the superfast propagation would not be possible
without the initial carriers generated by tunneling. Hence both
tunneling and impact ionization are essential for the propagation
mechanism. This type of an ionization front can be called
a {\it tunneling-assisted impact ionization} front.

As the front propagates, the current $I$ increases and the voltage on
the device $U$ decreases due to the interaction with the external load.
The maximum value of the electrical field and the front velocity
slightly increase during the passage (Fig.\ 3(b)).
As the front comes closer to the left boundary,
the field in the high-field region of the $n-$base increases and
the region of tunneling impact ionization becomes wider.
This effect also contributes to the acceleration of the front,
and causes the front interface to become smoother
(compare curves 1,2 and 3,4 in Fig.\ 3).
Eventually, the tunneling process becomes efficient
in the whole region between the moving front
and the $n^{+}-$contact. As a result, the front ceases
to propagate. Rather the rest of the $n-$base breaks down
by nearly uniform impact ionization.

The switching takes 30~ps
and fills the $n-$base
with electron-hole plasma of concentration $\sim 5 \cdot 10^{17}
\rm {cm^{-3}}$. The voltage drops from 8~kV to the residual
value of 10~V, applying an average
voltage ramp $\sim 300 \; {\rm kV/ns}$ to the load.
The falling part of the $U(t)-$dependence in Fig.\ 2
has a plateau that reflects the
transition from the formation of the initial nucleus of electron-hole
plasma to the stage of front propagation. During the front propagation stage,
the voltage drops from $\sim 7.4 \; {\rm kV}$ to 10~V within
less than 15~ps. Hence the "effective" voltage ramp is much higher
than the average one and reaches 500~kV/ns.

\section{The effect of random thermal generation}

The triggering
of the tunneling-assisted impact ionization front might fail due
to the thermal generation of free carriers in the depleted layer.
If during an early stage of development, thermal carriers lead
to an avalanche breakdown and to the formation of local conducting channels
between the $p^+-$ and $n^+-$layers, the conductance might increase
and prevent the voltage $U(t)$ from increasing further.
In this case, a front cannot start. Therefore the related
time scales and their temperature dependence have to be estimated.

The characteristic rate of thermal generation of electron-hole pairs
can be deduced from the value of the leakage current density.
It is $\approx 10^{-7} \; {\rm A/cm^2}$ at room temperature
\cite{GRE79,GRE89}.
Hence on average, one thermal electron-hole pair per ns is generated
in the whole volume of the $n-$base. Depending on the position
of the generation, the carriers need 0.5 to 1 ns to leave
the $n-$base of length 100 $\mu$m with saturated drift velocity
$10^7$~cm/s. If the field is still low, they will simply leave
the base, but if they are generated in a region with
$E > 2 \cdot 10^5 \; {\rm V/cm}$ or pass through such a region,
they will create an impact ionization avalanche on their path.
After 0.7 ns, the
tunneling-assisted impact ionization front leads to the breakdown
of the voltage. Though the probability that one ionization
avalanche will be initiated during the delay time is close to one,
this avalanche in most cases cannot form a conducting channel before
the front is triggered. The probability for more avalanches decreases
exponentially with their number.

However, even if a single conducting channel has developed, its cross-section
is too small compared to the cross-section of the system to influence the
triggering of the superfast front. Typical diameters of local
conducting channels created by solitary avalanches are
about $10 \; {\rm \mu m}$ \cite{microplasma}, so the cross-section
is about 1/1000 of the total cross-section of the semiconductor
structure.
Assuming concentrations of $n,p \sim 10^{18} \; {\rm cm^{-3}}$ and
high-field transport with saturated velocities $v_s$, we evaluate
the maximum current that can flow through such channel
as $\sim \;1$~A  \cite{MIN97}. This is comparable to the displacement current
in the structure during the delay stage, and far too small
to prevent the further increase of the voltage on a structure
with large cross-section.

The generation rate depends exponentially on the temperature \cite{Sze}:
\be
\nu(T) = \nu(T_0) \exp \left(\frac{E_g}{kT_0} -
\frac{E_g}{kT}\right),
\label{frequency}
\ee
where $k$ is Boltzmann's constant and $T_0$ is room temperature. 
For $T = 70 \; K$  and
$T = 400 \; K$, we find characteristic times of $\tau \sim 10 \; {\rm ms}$
and $\tau \sim 1 \; {\rm ps}$, respectively.
Hence the triggering mechanism is very sensitive
to the temperature of the structure: cooling the sample
is favorable whereas triggering of tunneling-assisted fronts
is likely to be impossible at high temperatures.

We conclude that the tunneling-assisted impact ionization front
can be triggered only, if the applied voltage increases sufficiently
fast and the threshold of tunneling ionization is reached faster than
1~ns. At room temperature and below, the triggering mechanism is
robust with respect to random initiation of impact ionization
avalanches by thermal carriers.
This conclusion is also supported by experimental
data for common impact ionization fronts\cite{GRE79,GRE89}:
for comparable experimental set-up and the lower voltage ramp of
$A \sim 1 \; {\rm kV/ns}$, the delay in deterministic triggering
of the impact ionization front could be as large as 3~ns.

\section{Limitations of the drift-diffusion model}

In this section, we discuss the applicability of the minimal
drift-diffusion model to the rapid high-field process described in
Sec.\ III.  Let us first briefly summarize the
characteristic scales of the obtained numerical solution:
the front length is $\ell_f \sim 3 \; {\rm \mu m}$,
the front velocity $v_f \sim 4 \cdot 10^8 \, {\rm cm/s}$,
the total switching time $\sim 30 \; {\rm ps}$,
the concentration behind the front $\sim 5 \cdot 10^{17} \rm {cm^{-3}}$,
and the current density $J \sim 10^5 \; {\rm A/cm^2}$.

{\it Relaxation time and electro-magnetic propagation time.}
A simple evaluation in the spirit of Drude theory and based on the
low-field mobility, estimates the upper bound for the electron
mean free path and the lower bound for the electron momentum
relaxation time as 20~nm and 0.2~ps, respectively. The
electro-magnetic propagation time is $W/c = 0.5 \; {\rm ps}$,
where $c$ is the speed of light. These scales are considerably
smaller than the respective scales of the process under study.

{\it Recombination.}
In Si for concentrations $n = p = 10^{18} \, {\rm cm^{-3}}$, the
inverse rate of Auger recombination is $\sim 10 \; {\rm \mu s}$;
the thermal recombination life-time is also of the order of microseconds
\cite{Landsberg}. Thus recombination can be neglected.

{\it Electron-hole scattering.}
For current densities above $10^2 \, {\rm A/cm^2}$,
electron-hole scattering which is not accounted
for in our model, becomes important \cite{MNA}.
Due to the electron-hole scattering, the resistivity $\rho$
of the dense plasma behind the front substantially decreases.
For $J \sim 10^3 \; {\rm A/cm^2}$ and $n,p \sim 10^{18} \; {\rm cm^{-3}}$
we estimate the order of the magnitude as
$\rho \sim 1 \; {\rm \Omega \cdot cm}$ \cite{MNA}. Hence the voltage drop
across the $n-$base after switching can be estimated
as  $J \varrho W \sim 10 \; V$.
This value should be added to the 10~V
obtained in our simulations. (Note that in our model
the residual voltage is mostly due to the recovery of
the electrical field on the $p^+-n-$junction
during the front propagation.) This difference of 10 V is negligible
compared to the kilovolt voltage drop during the front passage,
and does not change other characteristics of the switching process.

{\it Continuum approximation.}
The continuum treatment of electron and hole concentrations
is the major limitation of the standard drift--diffusion model
in application to the present problem.
Both tunneling and impact ionization
terms ignore the discrete nature of the ionization process:
The tunneling term predicts a small but steady
increase of the concentration even for electrical fields that
are far below the effective threshold of tunneling
ionization. In turn, the impact ionization generation term
(\ref{ion_coef}) models the multiplication of
any concentration of free carriers, even if these concentrations
physically correspond to a tiny fraction of an electron or hole
in the whole volume of the device.
Since, first, these small concentrations of initial carriers
are inherently present in the system at the early
stages of the process and, second, the system is very sensitive to
the appearance of free carriers in the high field region,
the unreflected use of an continuum
approximation leads to physically meaningless results:
triggering of the front is observed at very low electrical
fields due to the multiplication of unphysically small
concentrations of free carriers. Such an unphysical solution
that predicts premature triggering of the front
is depicted by the dashed line in Fig.\ 6:
the front is triggered  at $t \approx 300 \; {\rm ps}$
when the maximum electrical field is not more than
$E(W,t) = 5 \cdot 10^5 \; {\rm V/cm}$, which is too low for
tunneling ionization.

These unphysical solutions can be eliminated  by introducing
the cut-offs for low concentrations
in the impact ionization generation term
(\ref{impact_ionization}). The cut-off concentrations $n_{cut}$
and $p_{cut}$ approximate the range of validity of the continuum
approximation. Carefully chosen cut-offs allow to avoid
the artifacts and to provide qualitatively relevant
results. In Fig.\ 6, we show the transient characteristics for
different cut-offs $n_{cut}$. In a wide range of physically
meaningful parameter values of $n_{cut}$, the front triggering
and propagation is qualitatively the same.
However, the time delay
in the front triggering somewhat
depends on the cut-off level,
thus making the continuum model unsuitable for accurate
quantitative predictions.
The delay time shifts by approximately 50~ps when the cut-off
level changes by one order of magnitude in both directions,
whereas the triggering time remains approximately the same.
A quantitative analysis would demand
a more elaborate stochastic microscopic model.

\section{Discussion}

Tunneling-assisted impact ionization fronts are quite similar to
the well studied case of superfast impact ionization fronts
that underly the operation of TRAPPAT-diodes \cite{PRA68,DEL70}
and sharpening diodes \cite{GRE79,BEN85,GRE89,FOC97,KAR95}.
In both cases we deal with a collective phenomenon of superfast
propagation. It is based on avalanche multiplication of the already
existing carriers due to impact ionization in a finite narrow region
of the device, followed by screening of the electrical
field due to Maxwell relaxation in the adjacent spatial region.
The important difference is the source of the free carriers that
initiate the avalanche impact ionization process.
TRAPATT-like fronts propagate
into a depleted $n-$base with a certain concentration of initial
carriers, often referred to as "pre-ionization" \cite{DYA88,DYA89}.
The physical mechanism that creates "pre-ionization" then is
unrelated to the front propagation mechanism, e.g., in the case of
the TRAPPAT-diode, these are non-equilibrium carriers left behind
after the previous front passage.
In the case under consideration here, there is no pre-ionization
and the initial carriers are generated during the front passage
by tunneling in a region just ahead of the impact ionization front.
Hence tunneling and impact ionization coherently cooperate in the
superfast propagation of the tunneling-assisted impact ionization front.

For impact ionization fronts that propagate into homogeneously
pre-ionized media, an essential ingredient is the spatial profile
of the electrical field: the field should be below the threshold
of impact ionization
at a certain distance from the front \cite{DEL70,DYA88,DYA89}.
This keeps the "active" region
where the impact ionization develops, finite
and prevents a quasi-uniform blow-up of the concentration in the
whole sample. For ionization fronts in $p^+-n-n^+-$structures this
profile is due to the doping of the $n-$base that gives
a slope of the electrical field $q N_d /\varepsilon \varepsilon_0$
in the depleted layer. In contrast,
for the tunneling-assisted front, the electrical field is well above the
threshold of impact ionization in the whole $n-$base. The size
of the "active" region is controlled by the threshold of
tunneling ionization. This allows for much higher electrical fields,
increasing the front velocity and the concentration of the generated
plasma by orders of magnitude.

The possibility of tunneling ionization fronts has been discussed
before in Ref.\ \onlinecite{Konstantinov}. The theoretical
investigation in  Ref.\ \onlinecite{Konstantinov} takes only
tunneling ionization into account, assuming that in high fields
the impact ionization component can be neglected.
Our results show that this is not possible for electrical fields
of the order of $10^6 \; {\rm V/cm}$: the importance of impact
ionization increases with concentration and dominates tunneling
ionization already for concentrations of free carriers as low as
$n,p > 10^{13}\; {\rm cm^{-3}}$,
as can be read from Fig.\ 4.
Thus impact ionization is the dominant mechanism of free carrier
generation even in the range of fields where tunneling is possible,
after a sufficient carrier concentration is reached. Accordingly,
simulations with the impact ionization term set to zero show
that the tunneling ionization alone does not lead to front propagation;
rather the breakdown of the sample becomes quasi-uniform.
We suggest the following explanation for this observation:
the appearence of the traveling front solutions is known
to be due to an auto-catalytic dependence of the impact ionization
generation rate on the concentration of initial carriers
\cite{DEL70,DYA88,DYA89,EBE96}.
It leads to the exponentially fast increase of the concentration
with time in a given electrical field.
The rate of tunneling ionization lacks this auto-catalytic dependence
on the concentration, predicting an algebraically slow growth of
concentration with time.

\section{Summary}

We have described the theory of a new type of ionization fronts
in layered Si semiconductor structures. These are superfast
tunneling-assisted impact ionization fronts that propagate
due to the coherent effect of tunneling and impact ionization.
The front propagates into the fully depleted $n-$base of a
$p^+-n-n^+-$structure. The region of tunneling ionization
in electrical fields higher than $10^6 \; {\rm V/cm}$
has a characteristic length of dozens of micrometers. The generated
carriers initiate an avalanche impact
ionization process that becomes much more efficient than tunneling
as concentration increases. The concentration front
has a characteristic width of several micrometers.
Within  this length the concentration of free carriers increases
to the level of $n,p > 5 \cdot 10^{17} \; {\rm cm^{-3}}$.
Maxwell relaxation in the generated electron-hole
plasma leads to the full screening of the applied electrical field.
The front propagation is a collective phenomenon based on
ionization and screening, and its velocity $v_f$ is not limited by
the high-field drift velocity of individual carriers $v_s$. We observe
$v_f \approx 40 \cdot v_s$ in our simulations.

The front triggering becomes possible if the threshold of tunneling
ionization $10^6  \; {\rm V/cm}$ is reached after a delay shorter
than 1~ns. This ensures that random thermal ionization in the depleted
layer does not spoil the triggering. This condition is met at room
and lower temperatures when a voltage ramp $\sim 10 \; {\rm kV/ns}$
is applied to the structure connected in series with a load.
Such voltage pulses are state of the art in modern semiconductor
pulse power electronics \cite{GRE89}. The passage of the tunneling-assisted
impact ionization front switches the structure into the conducting state
with a residual voltage of several dozens of volts. The transient
$U(t)-$characteristic is nonlinear:
the total duration of the switching process observed is 30~ps,
whereas the effective switching time that corresponds to the
major voltage drop is below 15~ps.
Hence a voltage pulse of rise rate $\sim 500 \; {\rm kV/ns}$
and an amplitude of several kilovolts is applied to the load.
These values set new frontiers in pulse power electronics.

We finally remark, that the standard drift-diffusion model
has serious limitations when applied to models of superfast
ionization fronts. These limitations are due to the continuum approximation.
They manifest themselves in a premature unphysical triggering
of the front at low electrical fields due to avalanche processes
initiated by tiny fractions of electrons or holes. We have eliminated
these unphysical solutions by introducing a cut-off in the impact
ionization term. For further progress towards a fully quantitative 
description, microscopic stochastic model have to be investigated.

\acknowledgements

This work was supported by the Dutch physics funding agency FOM
and the program ``Generation of high power pulses of electrical
energy, particles and electro-magnetic radiation'' of the Presidium
of the Russian Academy of Sciences. P.R. also acknowledges support of
the Alexander von Humboldt Foundation.

\end{multicols}

\vspace{6cm}

\begin{figure}[tbp]
\setlength{\unitlength}{1cm}
\begin{center}
\begin{picture}(8,5)
\epsfxsize=8cm
\put(0,0){\epsffile{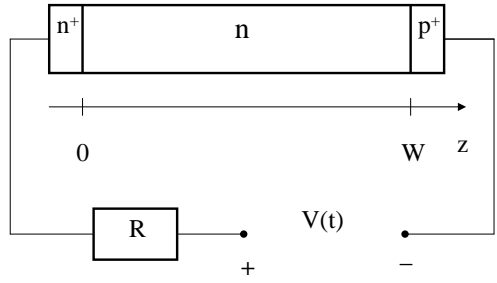}}
\end{picture}
  \end{center}
\label{sketch}
\caption{ Sketch of the  $p^{+}-n-n^{+}-$structure  operated in an external
circuit with load resistance $R$.
}
\end{figure}

\newpage

\begin{figure}[tbp]
\setlength{\unitlength}{1cm}
\begin{center}
\begin{picture}(10,18)
\epsfxsize=10cm
\put(0,0){\epsffile{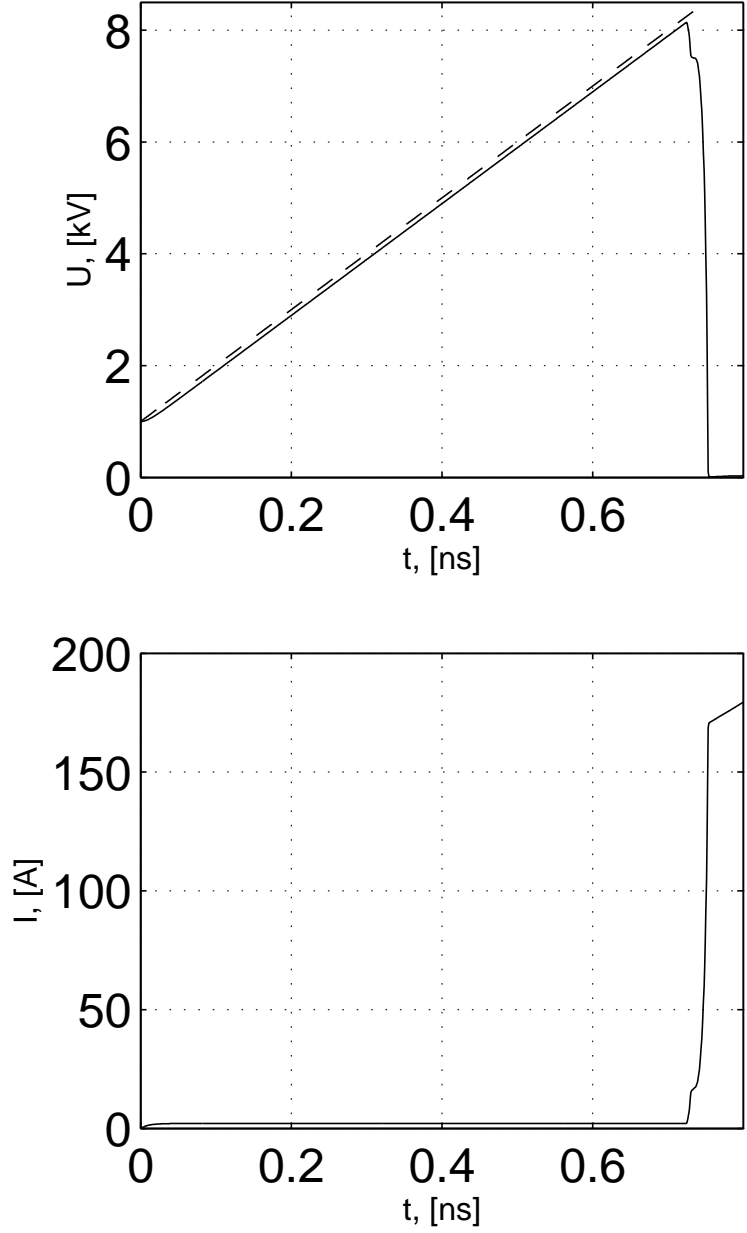}}
\end{picture}
  \end{center}
\label{U_I_t}
\caption{Voltage at the structure $U(t)$ (solid line in the upper panel)
and total current $I(t) = S \; J(t)$ (in the lower panel)
during the switching process.
The dashed line in the upper panel denotes the externally applied
voltage $V(t)$. The quantities shown are related through Ohm's law
$V = U + R I$.
Parameters: $W = 100 \; {\rm \mu m}$, $S = 0.002 \; {\rm cm^2}$,
$N_d = 10^{14} \; {\rm cm^{-3}}$, $N_a = 0$,
$V_0 = 1 \, {\rm kV}$, $A = 10 \; {\rm kV/ns}$,
$R = 50 \; {\rm \Omega}$,
$n_{cut} = p_{cut} = 10^{9} \; {\rm cm^{-3}}$.}
\end{figure}

\newpage

\begin{figure}
\setlength{\unitlength}{1cm}
\begin{center}
\begin{picture}(17,15)
\epsfxsize=16cm
\put(0,0){\epsffile{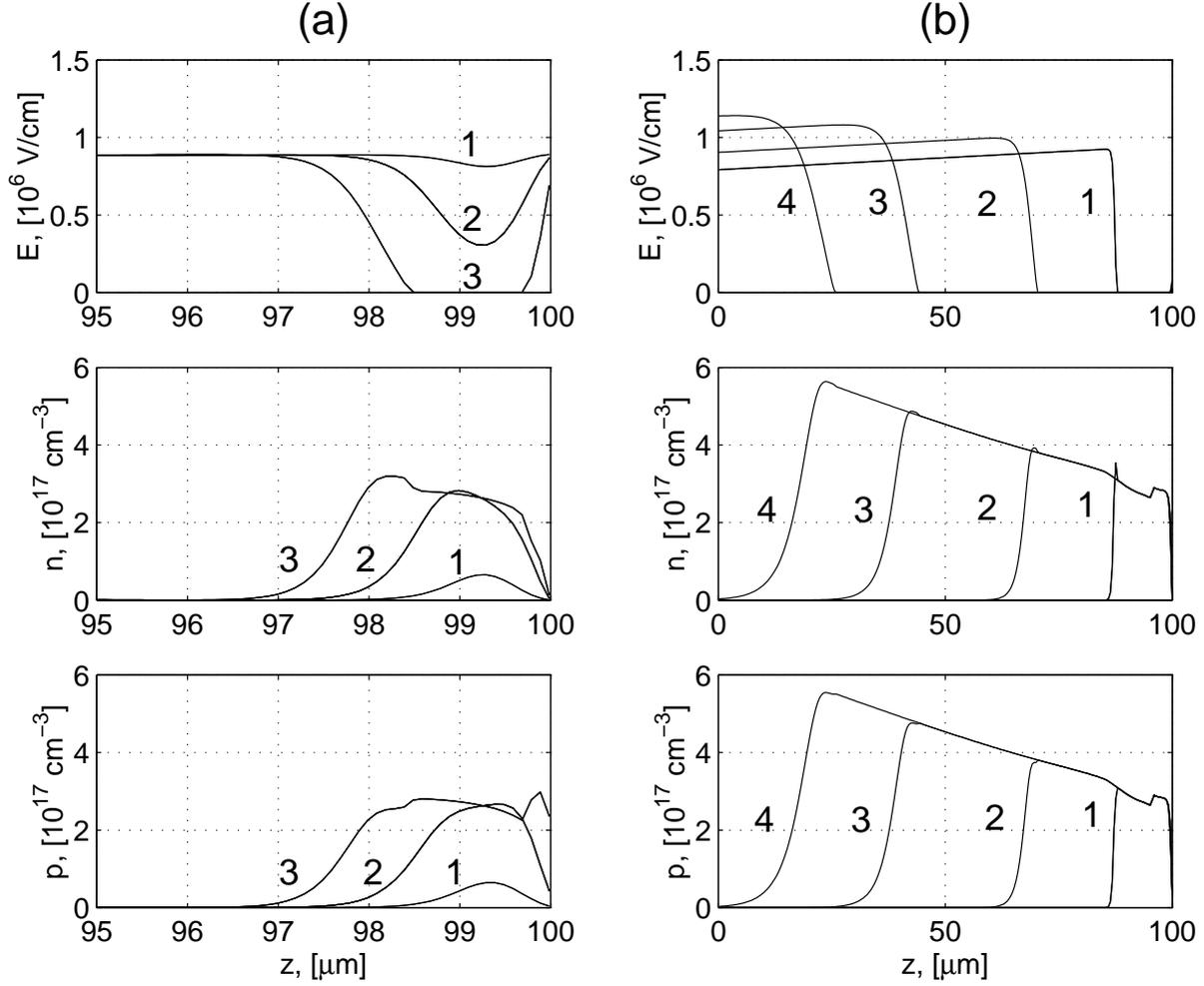}}
\end{picture}
\end{center}
%\vspace{2cm}
\label{fig3}
\caption{The internal dynamics leading to the external characteristics
of Fig.\ 2. Shown are the spatial profiles
of the electrical field $E(x,t)$ and electron
and hole concentrations $n(x,t)$, $p(x,t)$ in the $n-$base
($0 \le z \le W = 100 \, {\rm \mu m}$)
at different times:
(a) nucleation of electron-hole plasma and triggering of the
impact ionization front at times
$t = 725, \, 726, \, 727 \; {\rm ps}$
(curves 1,2,3);
(b) propagation of the tunneling-assisted ionization front at
$t = 735, \, 745, \, 750, \, 752 \, {\rm ps}$ (curves 1,2,3,4).
}
\end{figure}

\newpage

\begin{figure}
\setlength{\unitlength}{1cm}
\begin{center}
\begin{picture}(14,14)
\epsfxsize=14cm
\put(0,0){\epsffile{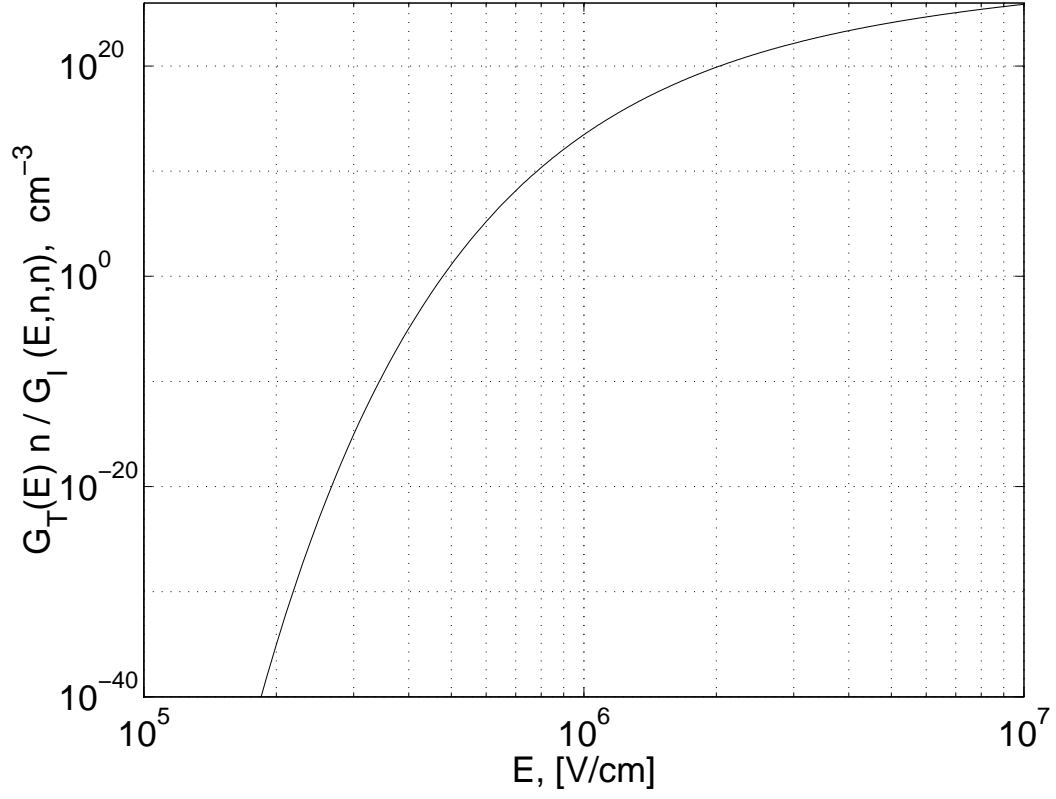}}
\end{picture}
\end{center}
\label{fig4}
\caption{Shown is the carrier concentration $n(E)$
for which impact ionization and tunneling ionization are
of equal strength. This concentration can be expressed as the ratio
$G_T(E) n /G_I (E, n, n)$, where we assumed $n=p$ and
neglected the cut-offs in (\ref{impact_ionization}).
}
\end{figure}

\newpage

\begin{figure}
\setlength{\unitlength}{1cm}
\begin{center}
\begin{picture}(10,17)
\epsfxsize=9cm
\put(0,0){\epsffile{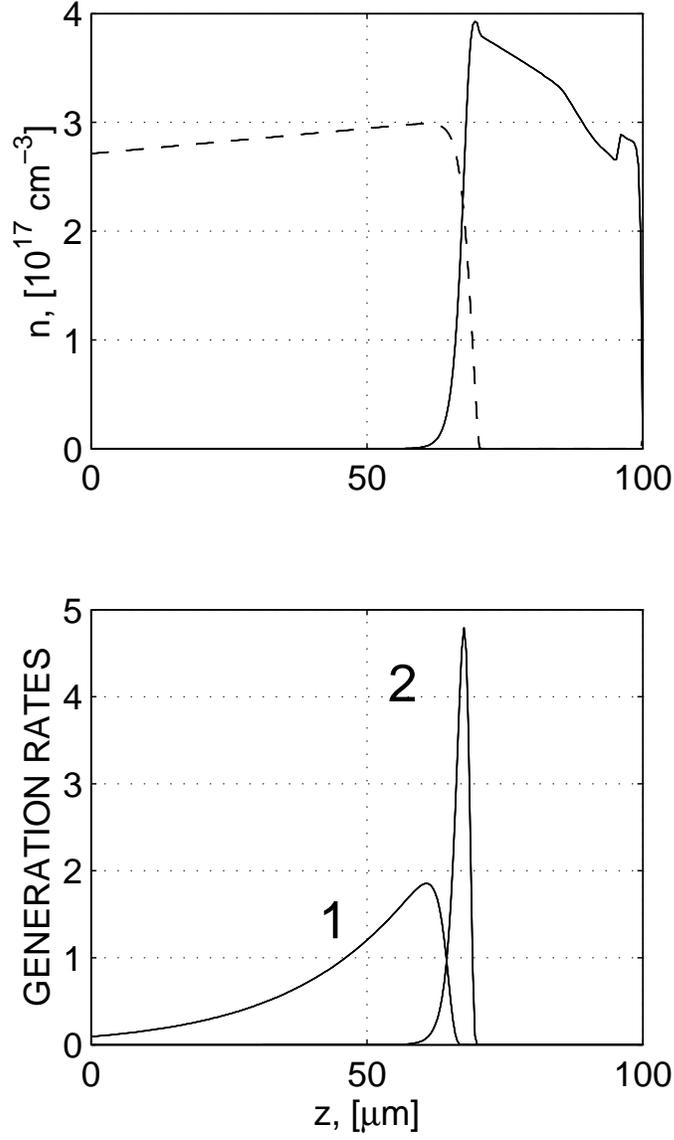}}
\end{picture}
\end{center}
\label{fig.5}
\caption
{The inner structure of the tunneling-assisted front
at some instant of time ($t = 740 \, {\rm ps}$).
The upper panel shows the electron concentration (solid line)
and the electrical field profile (dashed line, in arbitrary units).
The lower panel shows spatial profiles of the
tunneling generation rate $G_T$ (curve 1, one unit corresponds
to  $10^{27} \rm {/s \cdot cm^3}$ and the
impact ionization rate $G_I$ (curve 2, one unit corresponds
to $10^{36} \rm {/s \cdot cm^3}$).
Note the different scales of $G_T$ and $G_I$.}
\end{figure}

\newpage

\begin{figure}
\setlength{\unitlength}{1cm}
\begin{center}
\begin{picture}(12,12)
\epsfxsize10cm
\put(0,0){\epsffile{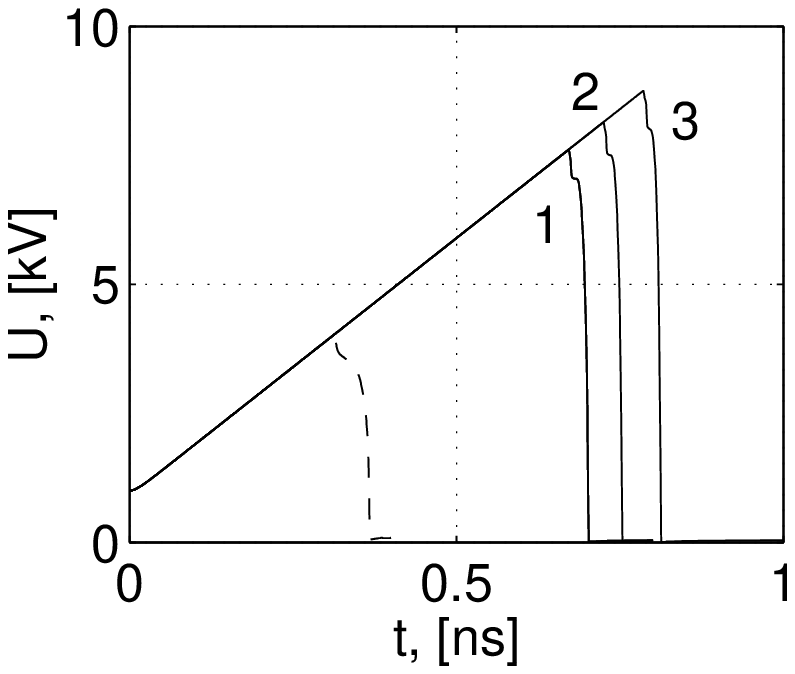}}
\end{picture}
\end{center}
\label{fig6}
\caption{Voltage at the device $U(t)$ calculated for different
cut-offs $n_{cut} = p_{cut} = 10^{8},\; 10^{9}, \;
5 \cdot 10^{9} \; {\rm cm^{-3}}$ (curves 1,2,3, respectively).
The dotted line shows the unphysical premature switching in
a low electrical field obtained for $n_{cut} = p_{cut} = 0$.
}
\end{figure}


\begin{references}
\bibitem{boundary}
Our results are not sensitive to boundary conditions.
In particular, the main characteristics of the process
remains the same, if homogeneous Neumann boundary conditions
are used for both electron and hole concentrations at both contacts.
\end{references}
\end{document}